\RequirePackage{fix-cm}
\documentclass[a4paper]{article}				% onecolumn (standard format)
\usepackage{a4wide}
\usepackage{graphicx}
\usepackage{amssymb}
\usepackage{hyperref}
\usepackage{natbib}
\usepackage{changebar}
\begin{document}

\title{Investigating sentence severity with judicial open data - \\
A case study on sentencing high-tech crime in the Dutch criminal justice system}

%\titlerunning{Short form of title}	% if too long for running head

% \author{}
\author{Pieter Hartel$^1$ \and
	Rolf van Wegberg$^1$ \and
	Mark van Staalduinen$^2$
}


% \institute{}
%% \institute{Pieter Hartel \at
%% 		TU Delft
%% 		\email{Pieter.Hartel@tudelft.nl}
%% 	\and
%% 		Rolf van Wegberg \at
%% 		TU Delft
%% 		\email{R.S.vanwegberg@tudelft.nl}
%% 	\and
%% 		Mark van Staalduinen \at
%% 		CFLW
%% 		\email{Mark.vanStaalduinen@cflw.com}
%% }

\date{%
    $^1$\{pieter.hartel,r.s.vanwegberg\}@tudelft.nl\\%
    $^2$mark.vanstaalduinen@cflw.com\\[2ex]%
    \today
}

\maketitle

\begin{abstract}
Open data promotes transparency and accountability as everyone can analyse it.
Law enforcement and the judiciary are increasingly making data available, to increase trust and confidence in the criminal justice system.
Due to privacy legislation, judicial open data -- like court judgments -- in Europe is usually anonymised.
Because this removes part of the information on for instance offenders, the question arises to what extent criminological research into sentencing can make use of anonymised open data. \\
 We answer this question based on a case study in which we use the open data of the Dutch criminal justice system that \href{https://www.rechtspraak.nl/Uitspraken}{rechtspraak.nl} makes available.
Over the period 2015-2020, we analysed sentencing in 25,366 court judgments and, in particular, investigated the relationship between sentence severity and the offender's use of advanced ICT -- as this is information that is readily available in open data. \\
 The most important results are, firstly, that offenders who use advanced ICT are sentenced to longer custodial sentences compared to other offenders.
Second, our results show that the quality of sentencing research with open data is comparable to the quality of sentencing research with judicial databases, which are not anonymised.
\end{abstract}
% \keywords{Focal concerns \and Cybercrime \and Sentence severity \and Open data \and Open source \and Coding software }

\section{Introduction}
\label{sec:introduction}
In the Netherlands, a judge has a great deal of freedom in sentencing decisions.
There is a universal minimum of 1 day in prison.
The maximums differ per offense, with a generic maximum of 30 years imprisonment and a life sentence for a handful of offenses.
Sentencing decisions are therefore mainly tailor-made.
The judge should impose the correct sentence, taking into account the personal circumstances of the offender, whilst ensuring the protection of society.
In the Netherlands, there are no strict guidelines, in contrast to the United States for example, wherein a judge is advised to use a narrower bandwidth for appropriate sentencing~\citep{Albonetti1997}.
However, there is a list of sentences that judges usually impose for the most common form of the criminal offense~\citep{LOVS2020}.
Judges can choose to consult that list, but the guidelines are not binding.
Hence, sentencing is a complex issue that is not necessarily addressed in the same way by all judges.

In theory, the variance in sentencing can be explained by the bounded rationality model~\citep{Simon1955}.
After all, the judge has to make a rational decision in a limited time and with limited information.
The judge is not a machine, and even experienced judges make quick, instinctive, and emotional decisions~\citep{Kahneman2011}.
The focal concerns theory~\citep{Hartley2014} tries to explain the variance in sentencing.
The theory states that sentencing is based on three aspects: blameworthiness, dangerousness, and practical considerations.
The type of crime and the suffering of the victims mainly determine blameworthiness.
The dangerousness is determined by the degree to which the judge believes that the offender will pose a threat to society after his release.
One of the main practical considerations is the limited capacity of the prisons.

Here, we apply the focal concerns theory to the Dutch penal system.
For example, a maximum sentence has been set for each offense, which is an important part of blameworthiness.
The legislator has also determined that the personal circumstances of the offender may influence the decision.
For example, an offender who has a job, a home, and a spouse has less chance of reoffending than an offender who is in less fortunate circumstances.
This aspect of sentencing is related to dangerousness.
Finally, criminal cases in the Netherlands are often settled with a fine, partly because of efficiency considerations.
This is an eminently practical consideration.

The legislator has also determined that certain aspects may not be taken into account -- e.g., sex, country of birth, and race.
In related research based on the focal concerns theory, attention has been paid to the extent to which such aspects have nevertheless been included in sentencing decisions.
One such aspect that is becoming increasingly more relevant in our modern world is the use of advanced Information and Communication Technology (ICT).
However, we have not been able to find related work on focal concerns theory where this aspect has been highlighted.
In modern society computers and networks cannot be ignored.
ICT also plays a role in crime~\citep{Raets2019}, and therefore in court cases~\citep{Bijlenga2018}.
Victims are threatened via social media~\citep{Montoya2013}, phishing is the modern modus operandi of fraud~\citep{Leukfeldt2014}, drug trafficking takes place through the TOR network~\citep{Christin2013}, and Bitcoins are used for extortion and ransomware~\citep{Connolly2019}.
It seems plausible that an offender who has used a certain technology also has to have some level of knowledge or skills to handle that technology.
For example, an offender who launders money with the use of Bitcoins is likely to know how to acquire Bitcoins, how they should be stored, and how one can launder them.
The offender does not need to have in-depth knowledge of the mathematical basis of Bitcoins, but he must have at least a Bitcoin wallet and be able to deal with it.

% \cbstart
We assume that not all judges know enough about ICT to make a rational decision on technically complex cases.
That is why we are specifically interested in the use of ICT by the offender and the response of the judicial system.
This makes it opportune to apply the focal concerns theory to the ICT knowledge or skills of both the offender and the prosecution.
A possible technical lack of knowledge on the part of the judge will probably have the most influence on the second focal concern, the dangerousness.
Could the offender be a technical expert who can be considered capable of realizing new threats?
Can an offender with better technical knowledge hide from the police for longer?
Could the offender even lead a criminal organization while incarcerated?
An expert can assist the court, but experts are scarce, and therefore practical considerations (third focal concern) are also relevant here.
As far as we know, the role of ICT in focal concerns theory has not been investigated before.
This is a gap in criminological research that we will address in a case study on sentencing high-tech crime in the Dutch criminal justice system (Sections \ref{sec:background} \dots \ref{sec:discussion}).
In the remainder of this section we establish a quality criterion for sentence severity research to be able to compare the case study to related work.
% \cbend

\begin{table}
\caption{Related work}
\label{tab:related_work}
\begin{tabular}{l @{} r l @{} r l @{} l @{} r r r }
\hline\noalign{\smallskip}
Reference			&N	&data	&N	&$R^2$	&		&mean	&mean	&N \\
			&citations	&from	&	&	&		&age	&months&cat. \\
\noalign{\smallskip}\hline\noalign{\smallskip}
\citet{Snodgrass2011}		&69	&NL	&4683	&0.01	&		&26	&	&1 \\
\citet{McAdams2016}		&14	&US	&434876	&0.02	&\checkmark	&23	&	&1 \\
\citet{Marcum2011}		&10	&US	&62	&0.09	&		&36	&18.0	&4 \\
\citet{Holtfreter2013}		&24	&US	&173	&0.11	&		&41	&41.4	&7 \\
\citet{Hartley2011a}		&16	&KR	&456	&0.15	&\checkmark	&35	&15.6	&9 \\
\citet{Patrick2011}		&31	&US	&42552	&0.20	&\checkmark	&33	&204.0	&8 \\
\citet{Lynch2018}		&13	&US	&918	&0.22	&		&	&	&6 \\
\citet{Lee2011}			&16	&KR	&1553	&0.22	&		&33	&11.8	&9 \\
\citet{Leifker2011}		&17	&US	&500	&0.23	&		&30	&4.5	&14 \\
\citet{Maddan2012}		&29	&US	&625	&0.24	&		&	&	&12 \\
\citet{Koons-Witt2014}		&56	&US	&22828	&0.31	&		&	&55.0	&6 \\
\citet{McEwen2015}		&30	&US	&202	&0.32	&		&28	&	&8 \\
\citet{Viglione2011}		&157	&US	&12103	&0.32	&		&	&	&6 \\
\citet{Doherty2016}		&10	&RW,YU	&131	&0.34	&\checkmark	&	&	&9 \\
\citet{Freiburger2010}		&53	&US	&2635	&0.35	&		&	&	&10 \\
\citet{Abrams2011}		&64	&US	&42552	&0.35	&\checkmark	&	&	&6 \\
\textbf{High-tech crime case study}&	&\textbf{NL}&\textbf{4125}&\textbf{0.39}&\textbf{\checkmark}&\textbf{38}&\textbf{24.8}&\textbf{8}\\
\citet{Peterson2013}		&70	&US	&531	&0.42	&		&	&	&10 \\
\citet{Huang2010}		&25	&US	&3709	&0.52	&\checkmark	&	&	&4 \\
\citet{Wermink2015}		&11	&NL	&17001	&0.56	&\checkmark	&	&6.7	&9 \\
\citet{Wermink2017}		&14	&NL	&1619	&0.59	&		&	&12.0	&12 \\
\citet{Wingerden2016}		&35	&NL	&9854	&0.60	&		&33	&11.0	&13 \\
\citet{Pina-Sanchez2014}	&22	&UK	&1949	&0.62	&		&	&	&7 \\
\citet{Rydberg2018}		&13	&US	&4562	&0.65	&		&33	&39.3	&10 \\
\citet{Hester2017}		&38	&US	&6611	&0.66	&\checkmark	&31	&16.0	&7 \\
\citet{Berdejo2013}		&125	&US	&18447	&0.73	&		&29	&67.0	&8 \\
\citet{Cohen2019}		&59	&US	&42552	&0.78	&		&36	&59.0	&11 \\
\citet{Feldmeyer2011}		&165	&US	&131672	&0.81	&		&33	&31.8	&12 \\
\citet{Spohn2014}		&23	&US	&2833	&0.86	&		&33	&72.4	&13 \\
\noalign{\smallskip}\hline
\end{tabular}
\end{table}

\subsection{Related work}
\label{subsec:related_work}
Earlier work that studied sentencing severity is based on the Anglo-Saxon legal system -- predominantly, the United States and the United Kingdom.
This has specific aspects, such as plea-bargaining, presumptive sentences, and mandatory minima that do not exist in the Dutch legal system.
As a result, American sentencing investigations cannot simply be compared with Dutch investigations.
However, studies from different jurisdictions that are based on focal concerns theory should be comparable.

It is common in sentencing research to predict the length of the custodial sentence with a linear regression on independent variables such as the age and sex of the offender, offense type, prior record, etc.
The best regression models explain the largest percentage of the variance in sentence length.
That is why we take the $R^2$ as a measure of the quality of the research.

To gain insight into the state-of-the-art in sentence severity research, Table~\ref{tab:related_work} provides an overview of articles published in the last 10 years, in which a linear regression model predicts the length of detention.
The articles included mainly differ in the choice of independent variables for the regression model and the population.

We first searched for journal articles with ``sentence length'' and $R^2$ in the full text using Google Scholar.
This resulted in 167 articles, of which those with at least 10 citations were analysed further.
We exclude articles that only mention the $R^2$ of a logistic regression, keeping only articles with a linear regression.
Finally, we identified 28 articles that report the explained variance in sentence length.
Of those 28 articles, we have coded the independent variables into 31 categories such as age, education, guideline departure, judge characteristics, mandatory minimum, etc.
The columns of Table~\ref{tab:related_work} show the reference, the number of citations on January 2021, the country from which the data originated, the N of the OLS model, the $R^2$ reported, a checkmark when the $R^2$ was adjusted, and the number of categories independent variables in the OLS model.
The table is sorted by the $R^2$, from lowest to highest.
Table~\ref{tab:related_work} shows that most of the literature relates to the American legal system.

\citet{Wermink2015} provide an overview of the Dutch criminal justice system and the extent to which the theory of focal concerns applies to it.
Subsequently, both \citet{Wingerden2016} and \citet{Wermink2017} study the influence of the personal circumstances of the offender on the sentencing.
In a recent article, \citet{Light2021} investigate the influence of nationality on sentencing.

The benchmark for Dutch sentencing research is set by \citet{Wingerden2016}, with an $R^2$ of 0.60.
Their research shows that the modus operandi has an influence, but the details of the modus operandi - such as the use of ICT by offenders - were not examined.
\citet{Wingerden2016} have also not paid attention to relevant knowledge about the modus operandi of the Public Prosecution Service and the Judiciary.
None of the articles mentioned in Table~\ref{tab:related_work} focus on ICT knowledge or skills of offenders and judges.

\citet{Hadzhidimova2019} have analysed press releases from the U.S. Department of Justice about $N=225$ cybercrime offenders who were not born in the US.
Their main conclusion was that sentencing severity depends on the type of cybercrime.
The influence of the application of advanced ICT on sentencing has therefore been studied to a limited extent, probably because this is a relatively new topic and because perhaps not all criminology researchers have sufficient affinity with ICT.

\subsection{Open data}
\label{subsec:open_data}
Making court decisions available as open data promotes trust and confidence in the judiciary~\citep{Bargh2017}.
Lawyers and journalists can do their job better if they have a complete overview of case law.
Offenders and victims have more insight into the cases in which they are involved or cases that are comparable.
Policymakers can see the effect of their policies on court judgments.
And researchers do not have to wait for lengthy application procedures and can respond flexibly to current developments.
We give three examples of judicial open data sets:
\begin{itemize}
\item In the Netherlands, approximately 5\% of court decisions are available as anonymised open data via \href{https://www.rechtspraak.nl/Uitspraken}{rechtspraak.nl}.
The selection criteria are also available on \href{https://www.rechtspraak.nl/Uitspraken/Paginas/Selectiecriteria.aspx}{rechtspraak.nl}.
The main criteria are that court judgments are published in case of
(a) a crime against someone's life, or
(b) the maximum sentence is 4 years or more, or
(c) the case is particularly relevant for the general public, specific professionals, or for case law.

\item In England, non-anonymised essential information from some of the court judgments can be found in the commercial database of \href{https://www.thelawpages.com}{thelawpages.com}.
It is unknown which fraction of court judgments are available on  \href{https://www.thelawpages.com}{thelawpages.com} and which sampling method is used~\citep{Pina-Sanchez2019a}.

\item In the interest of transparency and fairness of justice, the Supreme Court of the United States has ruled that everyone has the right to access criminal records.
The open data of \href{https://www.courtlistener.com}{courtlistener.com} consists of millions of rulings and verdicts~\citep{Cao2020}.
\end{itemize}

There are several challenges in creating and using judicial open data~\citep{Bargh2017}:
\begin{itemize}
\item Judicial open data is never complete.
To protect the privacy of the people involved only the essence of a judgment is usually published so that statistics give a incomplete picture.
In exceptional cases, a court judgment is not anonymous, such as cases with a political connotation (ECLI:NL:RBDHA:2016:15014).

\item Laws and regulations evolve, and current laws and regulations must be observed when interpreting court judgments.
If this interpretation were not done carefully, it would seem as if there were changes in trends.

\item Judicial data is produced by a variety of organizations, such as the courts, law enforcement agencies, and the probation service.
This is not conducive to interoperability.
\end{itemize}

In the remainder of this article, we apply the methods and techniques of classic sentencing severity research to open data.
The first research question\footnote{The second reseach question is part of the case study} we will answer is:

\begin{quote}
RQ1: To what extent are the intrinsic limitations of open data a barrier to obtaining good results?
\end{quote}

As a definition of quality, we use the $R^2$ of the linear regression between the independent variables and the length of the custodial sentence.
% \cbstart
To compare the results of the case study with the parameters of related research, we have already included the results of the case study in Table~\ref{tab:related_work}, entry \textbf{High-tech crime case study}.
% \cbend

After the Background, Method, Results, and Discussion of the case study, we conclude the article with Conclusions on the usefulness of open data for criminological research.
There are two appendices to the article.
In Appendix~\ref{app:coding} we discuss in some detail how we developed the software used to code the open data.
In Appendix~\ref{app:issues}, we present a list of errors that we found in the published court judgments during the development of the software.

\section{Case study: Background}
\label{sec:background}
% \cbstart
We define \textbf{high-tech crime} as a type of crime in which the offender has used advanced ICT, such as malware, BitCoins, or a PGP phone.
% \cbend
\citet{Bijlenga2018} discuss in detail five cases that meet this definition.
Crime where standard applications of ICT, such as email, online banking, and online shopping are used, are not classified as high-tech crime.
We call this form of crime with standard applications of ICT \textbf{low-tech crime}.

High-tech crime is a phenomenon that has developed over the past 10 years.
Offenders have increasingly made use of advanced techniques.
In 2012, for example, the first high-tech criminal case in the Netherlands was published on \href{https://www.rechtspraak.nl/Uitspraken}{rechtspraak.nl}, but in 2020 there were 243 published cases.

In the Netherlands, there are several articles in the Penal code that are indicative of cybercrime, such as 138ab (Hacking), 139d (Eavesdropping), and 350d (Misuse of passwords).
Unfortunately, these articles of law do not distinguish between standard ICT and advanced ICT.
That is why we have chosen to search the court judgments for words and concepts that are characteristic of high-tech crime.
% \cbstart
For example, is the word Bitcoin used in the court judgment? Or has the judiciary engaged a technical expert from the Netherlands Forensic Institute (\href{https://www.forensischinstituut.nl/}{NFI}), the Maastricht Forensic Institute (\href{https://www.tmfi.nl}{TMFI}), or the National technology laboratory (\href{https://www.tno.nl}{TNO})?
% \cbend

To understand and appreciate the technical aspects of high-tech criminal cases, the police and judicial authorities have also had to study ICT in recent years.
Where initially little knowledge of high-tech crime was available, specialist teams have now emerged, such as the Team High-Tech Crime (THTC) of the national police, and the cyber chamber of the court in The Hague.
In addition, every judge, public prosecutor or lawyer may have to deal with ICT-related crime on a daily basis.

When committing a high-tech offense, at least one of the offenders is assumed to have relevant ICT skills.
Police and judicial authorities will therefore also need relevant knowledge when dealing with such a high-tech case.
And technology will also play a role in the court decision.
For example, the judge may rule that the offender in a high-tech case poses a greater danger to society than an offender without advanced ICT skills.
Or a judge who is not sufficiently familiar with the technical aspects of the case could be more cautious than usual.
The judge's considerations about the dangerousness of the offender may therefore be related to the technology.

One of the uses of ICT is in automating manual processes. 
Since crime is a process, it can be automated as well.
For example, with little effort, an offender may be able to reach millions of victims with a large-scale phishing campaign.
ICT can increase the scale of crime, and thereby increase the blameworthiness of the offender.
An offender with ICT skills may also be percieved as more dangerous.

Based on the focal concerns theory, we pose a second research question:

\begin{quote}
RQ2: To what extent do offenders in high-tech crime cases receive a more severe sentence than offenders in low-tech cases?
\end{quote}

A court judgment is usually accompanied by a press release and the indictment.
The court judgments that are available as open data on \href{https://www.rechtspraak.nl/Uitspraken}{rechtspraak.nl} were anonymised to guarantee the privacy of the offender.
Names, residential, IP and web addresses are replaced by indications such as \textit{[name 1]}, \textit{[co-offender 2]}, or \textit{[web address 3]}.
Some courts even omit the offender's year of birth.
Police case files, reports from the probation service, and other judicial documentation are never part of the open data.
There is therefore a large difference in the amount of detail that researchers using open data have compared to researchers who have access to complete case files.
This makes it challenging to develop high quality regression models.

%% Update when the WODC C%R report on 2020 (table 6.1) has appeared, and Exclude minors!
\begin{table}
\caption{Total number of criminal cases processed by Dutch district courts (N $\approx$ 520,000).
	(At the time of writing, data from the national statistics of 2020 was not yet available)}.
\label{tab:cases}
\begin{tabular}{l @{} *{6}{r} | r r }
\hline\noalign{\smallskip}
		&2015	&2016	&2017	&2018	&2019	&2020	&Total	&Percent \\
\noalign{\smallskip}\hline\noalign{\smallskip}
Adult men	& 86,640&77,225	&75,975	&72,530	&70,355	& 	&382,725& 86.6 \% \\
Adult women	& 13,010&11,350	&11,015	& 9,975	& 9,375	& 	& 54,725& 12.4 \% \\
Other		&    955&   925	&   875	&   840	&   925	& 	&  4,520&  1.0 \% \\
\noalign{\smallskip}\hline\noalign{\smallskip}
Total		&100,605&89,500	&87,865	&83,345	&80,655	& 	&441,970&100.0 \% \\
\noalign{\smallskip}\hline
\end{tabular}
\end{table}

\section{Case study: Method}
\label{sec:method}
Table~\ref{tab:cases} shows the total number of criminal court judgments in the years 2015 to 2020~\citep[Table 6.1]{Meijer2020}.
The total number of court judgments in the period 2015 to 2020 is estimated to be 520,000.
In the same period, 49,790 criminal court judgments were published as open data by the Dutch district courts.
Of these court judgments, 23,186 contained only meta-data, and 1,238 concern minors.
Because meta-data does not contain sufficient information for sentencing research, we disregard incomplete court judgments.
We also omit all court judgments where the offender is minor because they are sentenced under a different regime than adults.
The total number of complete court judgments with an adult offender is 25,366, or about 5\% of the national total.

We analyse the obtained criminal court judgments in two ways:
\begin{enumerate}
\item A linear regression is used to find the relationship between the sentence length and relevant information about the offender, the offence, and the judiciary.

\item Cohen's Kappa is calculated to investigate the reliability of the coding.
\end{enumerate}

\subsection{The Coding}
\label{subsec:coding}
A court judgment is a standardized document according to the PROMIS model~\citep{Hoven2011}, in which not only the decision but also the relevant details of the offender, the offence, and the police investigation are presented.
Normally the most important facts from the police reports are included in the court judgment.

Coding thousands of court judgments is time-consuming.
But because a court judgment has a standard structure, it is in principle possible to code a court judgment with an appropriate program.
In Appendix~\ref{app:coding} we describe in some detail how the program we developed codes the court judgments.
During the development of the program, we discovered several problems in the public data of \href{https://www.rechtspraak.nl/Uitspraken}{rechtspraak.nl}, such as:
\begin{itemize}
\item We have found court judgments that are not completely anonymised.
This has been reported to the relevant courts.

\item The legal basis is unclear in some court judgments.
For example, the court judgment ECLI:NL:RBROT:2018:3004 mentions a non-existent article (10310) in the legal regulations: \textit{Having regard to articles 14a, 14b, ,, 36f, 57, 285, 10310, 311 312 and 417bis of the Penal code and Article 11 of the Opium Act} and in the court judgment ECLI:NL:RBLIM:2018:2963 the legal basis is missing.

\item The amount of fines, and the length of time periods are sometimes shown in numbers, sometimes in words, and sometimes as both.
There are court judgments where the words and numbers are inconsistent.
For example, \textit{sentences the offender to a term of imprisonment of 35 (thirty) months} (ECLI:NL:RBROT\discretionary{:}{}{:}2018:1383), and
\textit{sentences the defendant to community service, consisting of performing unpaid work for the duration of 160 (one hundred and eighty) hours} (ECLI:NL:RBOVE\discretionary{:}{}{:}2019:1363).
This has been reported to the relevant courts.

\item There are spelling mistakes in the court judgments, which the automatic coding takes into account.
For example, \textit{vijfig}, \textit{vijendertig}, and \textit{vijvenvijftig}.
Spelling mistakes have not been reported to the courts.
\end{itemize}

\subsection{The data}
\label{subsec:data}
The program coded the sentencing decision of 24,994 (98.6\%) court judgments.
The distribution of the type of sentence is custodial sentence: 54.8\%, community service: 10.2\%, fine: 5.8\%, acquittal: 10.2\%, procedural court judgments (such as a ruling on extradition): 14.7\%.
In 1.4\% the sentencing decision could not be automatically coded.
Manual analysis of a random sample of court judgments that could not be automatically coded has shown that these are mainly procedural court judgments.

For 20,227 (79.7\%) court judgments, the program was able to code the offenses for which the offender was convicted.
Of the remaining 20.3\% court judgments, the coding program did not find the legal basis (\textit{Wettelijke voorschriften}), mainly because these are usually not mentioned if the offender is acquitted, or in procedural cases.

The top half of Table~\ref{tab:offences} presents the national statistics of court judgments for adults~\citep[Tables 6.2 and 6.12]{Meijer2020}.
The bottom half of Table~\ref{tab:offences} shows the statistics of the court judgments for adults from the open data set.
All offences are classified using the standard crime classification of Statistics Netherlands~\href{https://www.cbs.nl/nl-nl/onze-diensten/methoden/classificaties/misdrijven/standaardclassificatie-misdrijven-2010}{cbs.nl}, where the offence with the most severe sentence determines the classification of a judgment.
The open dataset contains considerably more violent crime (31.6\%) than the national data (9.7\%), but this is consistent with the publication criteria of the open data (see Section~\ref{subsec:open_data}).

\begin{table}
\caption{A comparison of the main type of offence per year.
	(At the time of writing, data from the national statistics of 2020 was not yet available)}.
\label{tab:offences}
	\begin{tabular}{l @{} *{6}{r} | r r }
\hline\noalign{\smallskip}
Offence type		&2015	&2016	&2017	&2018	&2019	&2020	&Total	& Percent \\
\noalign{\smallskip}\hline\noalign{\smallskip}
\multicolumn{9}{c}{National statistics for adults (N $\approx$ 520,000)} \\
\noalign{\smallskip}\hline\noalign{\smallskip}
Property		& 36,590&34,760	&32,970	&30,275	&29,405	&	&164,000& 37.1\% \\
Violent			&  9,835& 9,115	& 8,595	& 7,825	& 7,345	&	& 42,715&  9.7\% \\
Public order		& 20,250&18,405	&16,830	&15,440	&14,565	&	& 85,490& 19.3\% \\
Other Penal code	&  2,670& 2,490	& 2,040	& 1,840	& 1,750	&	& 10,790&  2.4\% \\
Road traffic		& 17,950&12,480	&14,115	&14,980	&15,905	&	& 75,430& 17.1\% \\
Drugs related		&  6,580& 6,685	& 6,495	& 6,170	& 5,675	&	& 31,605&  7.2\% \\
Weapons related		&  1,335& 1,195	& 1,050	& 1,035	& 1,000	&	&  5,615&  1.3\% \\
Other criminal		&  5,395& 4,365	& 5,745	& 5,775	& 5,005	&	& 26,285&  5.9\% \\
\noalign{\smallskip}\hline\noalign{\smallskip}
Total			&100,605&89,495	&87,840	&83,340	&80,650	&	&441,930&100.0\% \\
\noalign{\smallskip}\hline\noalign{\smallskip}
\multicolumn{9}{c}{Open data statistics for adults (N=20,227)} \\
\noalign{\smallskip}\hline\noalign{\smallskip}
Property		&  801	&  675	&  996	&1,101	&1,101	&  975	& 5,649	& 27.9\% \\
Violent			&  912	&  765	&1,106	&1,255	&1,214	&1,131	& 6,383	& 31.6\% \\
Public order		&  228	&  249	&  388	&  388	&  370	&  342	& 1,965	&  9.7\% \\
Other Penal code	&   81	&  127	&  131	&  141	&  164	&  150	&   794	&  3.9\% \\
Road traffic		&  147	&   90	&  153	&  158	&  147	&  180	&   875	&  4.3\% \\
Drugs related		&  304	&  265	&  405	&  444	&  518	&  515	& 2,451	& 12.1\% \\
Weapons related		&   81	&   58	&   93	&  110	&  149	&  132	&   623	&  3.1\% \\
Other criminal		&  175	&  187	&  264	&  281	&  333	&  247	& 1,487	&  7.4\% \\
\noalign{\smallskip}\hline\noalign{\smallskip}
Total			&2,729	&2,416	&3,536	&3,878	&3,996	&3,672	&20,227	&100.0\% \\
\noalign{\smallskip}\hline
\end{tabular}
\end{table}

\subsection{Hierarchical linear modelling (HLM)}
\label{subsec:hlm}
We ran a hierarchical linear regression to predict the length of the incarceration.
The hierarchy consists of three levels:
\begin{enumerate}
\item The first model is the baseline that contains only the control variables, such as the maximum sentence for the offence, and the dummy-coded set of offenses for which the offender is convicted.
This level corresponds to the first focal concern, blameworthiness.

\item The second model also contains the variables that relate to prosecution, such as whether sentencing guidelines were consulted, and whether the judiciary has enlisted the services of technical experts.
We propose to expand the third focal concern, the practical matters, with the technical knowledge of the judiciary.

\item The third model contains the variables related to the offender, such as sex, country of birth, age, prior record, and technical skills.
This level corresponds to the second focal concern, the dangerousness.
\end{enumerate}

We use factor analysis to determine whether there are latent variables that better explain the variance in sentencing than the manifest variables.
We also look for interaction effects, to rule out that certain independent variables influence the effect of other independent variables on sentencing~\citep{Gottschalk2014,Maddan2012,Ulmer2004,Wu2012}.

\subsection{Variables}
\label{subsec:variables}
The program automatically coded the variables listed below for the 25,366 court judgments.
We first list three categories of independent variables, followed by the dependent variable.

\paragraph{Offence related independent variables}
\begin{itemize}
\item \textit{Property offense}, \textit{Violent offense}, \textit{Public order offense}, \textit{Other Penal code offence}, \textit{Road traffic offence}, \textit{Drugs related offence}, \textit{Weapons related offence}, and \textit{Other criminal offense}: (0 = no, 1 = yes)
These seven variables represent the offense type, according to the standard crime categories of Statistics Netherlands~\href{https://www.cbs.nl/nl-nl/onze-diensten/methoden/classificaties/misdrijven/standaardclassificatie-misdrijven-2010}{cbs.nl}.
Since an offender can be convicted of more than one offence, we have dummy-coded the offence with the longest maximum sentence.

\item \textit{Number of offences}: (0 to 8)
This variable counts the number of offences.

\item \textit{Maximum prison months}
We have calculated the maximum length of imprisonment by looking up the relevant articles in the Penal Code as mentioned in the legal basis in the court judgment.
With one exception, we have omitted the articles from Book One of the Penal Code because these articles typically indicate the framework within which the sentences in Book Two must be interpreted.
The exception is that in some cases no other articles are mentioned than from Book One, for example, money-laundering convictions based on article 36f Book One, Penal code.
In that case, we took the maximum from Book One.
We have not differentiated between different custodial sentences, such as involuntary commitment, imprisonment, and military detention.
For life sentences, we have taken the legal maximum of 360 months.
Like age, this variable is not normally distributed.
Therefore, we dummy-coded the maximum -- in months -- as follows:
\textit{71 and shorter}, \textit{72-95}, \textit{96-107}, \textit{108-119} (reference), \textit{120-143}, \textit{144-179}, \textit{180-215}, and \textit{216 and longer}.
These categories were chosen to achieve an optimal balance in the number of cases in each category.
The smallest category \textit{216 and longer} holds 9.2\% of the cases, and the largest category \textit{144-179} has 16.6\% of the cases.
\end{itemize}

\paragraph{Prosecution related independent variables}
\begin{itemize}
\item \textit{Have guidelines been mentioned} (0 = no, 1 = yes)
If the judgment refers to the LOVS guidelines~\citep{LOVS2020}, or specific guidelines of the Public Prosecution Service, we assume that the court has taken this into account.
Otherwise, we assume that no guidelines were consulted.
So there are no missing items.

\item \textit{Has prosecution special expertise} (0 = no, 1 = yes)
To determine whether the police and the judiciary had deployed special expertise, we checked whether organizations such as NFI, TMFI, or TNO were mentioned in the court judgment or whether an expert witness was mentioned in the court judgment.
We have not taken into account the area of expertise of the specialists, which might include ICT, but also firearms, DNA, and drugs analysis.
In the absence of such a mention, we assume that no special expertise has been deployed.
So there are no missing items.
\end{itemize}

\paragraph{Offender related independent variables}
\begin{itemize}
\item \textit{Age at sentencing}
The age is the difference between the year of birth and the year the sentence was pronounced.
In some cases, this is a few years later than the age at the time of the offense because the judicial process sometimes takes years.
The year of birth is often anonymised in open data so that there are relatively many missing items.
This variable is not normally distributed.
To be able to compare the results, we dummy-coded the age into the same five categories as \citet{Wingerden2016}:
\textit{18-20}, \textit{21-30} (reference), \textit{31-40}, \textit{41-50}, and \textit{51 and older}.

\item \textit{Is born abroad} (0 = no, born in the Netherlands, 1 = yes, born abroad)
The offender's country of birth is usually stated explicitly in the court judgment.
We, therefore, assume that the offender was born in the Netherlands unless it is explicitly stated that he/she was born elsewhere.
Some court judgments do not mention where the offender was born.
In that case, the information is deemed missing.

\item \textit{Is female offender} (0 = no, male, 1 = yes, female)
The sex of the offender is not explicitly stated in the court judgment, but sometimes it is stated in the accompanying press release.
If the court judgment or press release contains fragments such as \textit{her co-offender}, then the offender is deemed to be female, otherwise, we assume it is a man.
This is consistent with the observation that most offenders are male.
There are no missing items.

\item \textit{Is repeat offender} (0 = no, 1 = yes)
Some court judgments state explicitly whether the offender has been in contact with the police and the judicial authorities before.
Often the court judgment only states that the court has looked at the judicial documentation, without making explicit whether the offender is a repeat offender or not.
In the latter case, the information is missing.

\item \textit{Has multiple victims} (0 = no, 1 = yes)
As with the offenders, the victims are anonymised as e.g. \textit{[victim 3]}.
From this, it can be deduced how many victims there were.
Without an indication from which the number of victims can be derived, the information is missing.

\item \textit{Has offender basic skills} (0 = no, 1 = yes)
The coding program searched the court judgments for ICT related standard terms such as \textit{Internet}, \textit{WWW}, \textit{online}, etc.
When we find one or more of those words, we assume that the offender has some basic level of cyber skills.
If we did not find any of those words, we assume that the offender has no demonstrable cyber skills.
So there are no missing items.

\item \textit{Has offender special skills} (0 = no, 1 = yes)
We also looked for specialist concepts such as \textit{Bitcoin}, \textit{Tor network}, and \textit{Ransomware} as evidence that the offender has more than just basic skills.
If we cannot find any of these words, the offender had no demonstrable special skills.
So there are no missing items.
To avoid correlation between the independent variables, we set \textit{Has offender basic skills} to 0 if there is special skills.
\end{itemize}

\paragraph{Dependent variable}
\begin{itemize}
\item \textit{Prison months} (logged)
To determine the length of the prison sentence imposed, we looked for textual passages in the court judgment in which the offender is convicted.
We did not take into account deductions for pre-trial custody or suspended sentences, and in the same way, as for the maximum we converted everything into months (= 30 days).
This variable is also not normally distributed, but after a log transformation, a visual inspection of the Normal Q-Q plots showed that the transformed variable is distributed normally.
\end{itemize}

\subsection{Reliability}
\label{subsec:reliability}
To check the reliability of the automatic coding, we manually checked the coding in a stratified random sample of 275 court judgments.
That is about 1\% of all complete court judgments.
Because courts do not necessarily record court judgments in the same way - for example, the Rotterdam District Court started an experiment in 2019 to make some court judgments more accessible - the 11 district courts form the first stratum.
We also expect the formulation to evolve, and therefore take the years 2015-2020 as the second stratum.
The stratification ensures that the distribution of the courts and the year of pronouncing the sentence in the random sample is the same as in the total collection of court judgments.
We calculate both the accuracy of the coding, as usual in the technical sciences, as well as Cohen's Kappa, which is more common in the social sciences.
We then manually checked coded information of all the court judgments in the sample: the age of the offender, the decision, and the legal basis of the decision.
\begin{itemize}
\item A true positive is when the coding program found the correct coding.
\item A false positive arises if the coding by the program is wrong.
\item If the program does not find any coding, while the relevant information is present, a false negative occurs.
\item If the information is missing, and the coding program noted this, we have a true negative.
\end{itemize}

\section{Case study: Results}
\label{sec:results}
We have found indications that, on average, high-tech crime leads to longer sentences.
The automatic coding is reliable.

\begin{table}
\caption{Descriptive statistics of ratio-scale variables}
\label{tab:scale_variables}
\begin{tabular}{l *{5}{r} }
\hline\noalign{\smallskip}
Scale variable			&N	&Minimum&Maximum&Mean	&Std. Deviation \\
\noalign{\smallskip}\hline\noalign{\smallskip}
Age at sentencing		&16772	&18	& 90	& 37.91	&13.23 \\
Number of offences		&20227	& 1	&  6	&  1.41	& 0.70 \\
Maximum prison months		&19527	& 0.47	&360	&129.98	&79.07 \\
Prison months			&14406	& 0.03	&360	& 24.79	&36.03 \\
\noalign{\smallskip}\hline
\end{tabular}
\end{table}

\begin{table}
\caption{Descriptive statistics dichotomous variables (N=25,366)}
\label{tab:dichotomous_variables}
\begin{tabular}{l *{7}{r} }
\hline\noalign{\smallskip}
Dichotomous variable		&Frequency &Percent \\
\noalign{\smallskip}\hline\noalign{\smallskip}
\multicolumn{2}{l}{Maximum prison months} \\
~~~~ $\leq$71				& 2,592	&10.2\% \\
~~~~ 72-95				& 2,809	&11.1\% \\
~~~~ 96-107				& 2,229	& 8.8\% \\
~~~~ 108-119				& 2,119	& 8.4\% \\
~~~~ 120-143				& 1,966	& 7.8\% \\
~~~~ 144-179				& 3,242	&12.8\% \\
~~~~ 180-215				& 2,772	&10.9\% \\
~~~~ $\geq$216				& 1,798	& 7.1\% \\
\multicolumn{3}{l}{Type of offence} \\
~~~~ Property offence			& 5,649	&22.3\% \\
~~~~ Violent offence			& 6,383	&25.2\% \\
~~~~ Public order offence		& 1,965	& 7.7\% \\
~~~~ Other Penal code offence		&   794	& 3.1\% \\
~~~~ Road traffic offence		&   875	& 3.4\% \\
~~~~ Drugs related offence		& 2,451	& 9.7\% \\
~~~~ Weapons related offence		&   623	& 2.5\% \\
~~~~ Other criminal offence		& 1,487	& 5.9\% \\
Number of offences			&20,227 &79.7\% \\
Have guidelines been mentioned		& 6,090	&24.0\% \\
Has prosecution special expertise	& 7,429	&29.3\% \\
\multicolumn{2}{l}{Age at sentencing} \\
~~~~ 18-20				&   779	& 3.1\% \\
~~~~ 21-30				& 5,249	&20.7\% \\
~~~~ 31-40				& 4,288	&16.9\% \\
~~~~ 41-50				& 3,250	&12.8\% \\
~~~~ $\geq$51				& 3,206	&12.6\% \\
Is born abroad				& 4,749	&18.7\% \\
Is female offender			& 2,042	& 8.1\% \\
Is repeat offender			& 7,238	&28.5\% \\
Has multiple victims			& 5,917	&23.3\% \\
Has offender basic skills		&12,160	&47.9\% \\
Has offender special skills		&   911	& 3.6\% \\
\noalign{\smallskip}\hline
\end{tabular}
\end{table}

\subsection{Descriptive statistics}
\label{subsec:descriptive_statistics}
Tables~\ref{tab:scale_variables} and~\ref{tab:dichotomous_variables} give the descriptive statistics of all variables.
All dependent variables except one are more or less constant over time.
The exception is the percentage of offenders with special ICT skills which, as can be expected, is steadily growing, from 2015: 1.6\%, 2016: 2.4\%, 2017: 2.9\%, 2018: 3.9\%, to 2019: 4.6\%, 2020: 5.2\%.

% =CONCATENATE(ROUND(D6132,2),IF(H6132<0.001,"**",""),IF(H6132<0.01,"*",""),IF(H6132<0.05,"*",""),"&",ROUND(E6132,2))

\begin{table}
\caption{Unstandardised coefficients of the three models for LN( sentence length )}
\label{tab:regression}
{\renewcommand{\baselinestretch}{1.4} 
\begin{tabular}{l @{}*{3}{r @{}l r} }
\hline\noalign{\smallskip}
				& \multicolumn{3}{c}{Model 1}	& \multicolumn{3}{c}{Model 2}	& \multicolumn{3}{c}{Model 3} \\
			& \multicolumn{3}{c}{(offence)}	& \multicolumn{3}{c}{(prosecution)}	& \multicolumn{3}{c}{(offender)} \\
Independent variable		& B	&	& S.E.		& B	&	& S.E.		& B	&	& S.E. \\
\noalign{\smallskip}\hline\noalign{\smallskip}
(Constant)			& 2.05	&***	&0.08		& 1.83	&***	& 0.07		& 1.49	&***	&0.08 \\
\multicolumn{10}{l}{Maximum prison months (108-119 ref.)} \\
~~~~ $\leq$ 71			&-0.70	&***	&0.08		&-0.63	&***	& 0.08		&-0.75	&***	&0.08 \\
~~~~ 72-95			&-0.80	&***	&0.08		&-0.71	&***	& 0.08		&-0.82	&***	&0.08 \\
~~~~ 96-107			&-0.25	&**	&0.09		&-0.29	&**	& 0.09		&-0.31	&***	&0.09 \\
~~~~ 120-143			&	&n.s.	&		&	&n.s.	&		&	&n.s.	& \\
~~~~ 144-179			& 0.51	&***	&0.09		& 0.46	&***	& 0.09		& 0.44	&***	&0.09 \\
~~~~ 180-215			& 1.12	&***	&0.06		& 1.05	&***	& 0.06		& 1.08	&***	&0.06 \\
~~~~ $\geq$216			& 1.35	&***	&0.08		& 1.20	&***	& 0.08		& 1.10	&***	&0.08 \\
\multicolumn{10}{l}{Type of offence (Violent ref.)} \\
~~~~ Property offence		&	&n.s.	&		& 	&n.s.	& 		&-0.12	&*	&0.05 \\
~~~~ Public order offence	&-0.81	&***	&0.07		&-0.76	&***	& 0.07		&-0.77	&***	&0.07 \\
~~~~ Other Penal code offence	&-0.58	&*	&0.26		&-0.50	&*	& 0.25		&-0.53	&*	&0.25 \\
~~~~ Road traffic offence	&	&n.s.	&		&-0.32	&**	& 0.12		&-0.29	&*	&0.12 \\
~~~~ Drugs related offence	&-0.62	&***	&0.12		&-0.77	&***	& 0.11		&-0.75	&***	&0.11 \\
~~~~ Weapons related offence	&	&n.s.	&		& 	&n.s.	& 		& 	&n.s.	& \\
~~~~ Other criminal offence	&	&n.s.	&		&	&n.s.	&		&	&n.s.	& \\
Number of offences		&0.17	&***	&0.02		& 0.15	&***	& 0.02		& 0.10	&***	&0.02 \\
Have guidelines been mentioned	&	&	&		& 0.08	&*	& 0.04		& 0.11	&**	&0.04 \\
Has prosecution special expertise&	&	&		& 0.63	&***	& 0.03		& 0.58	&***	&0.04 \\
\multicolumn{10}{l}{Age at sentencing (21-30 ref.)} \\
~~~~ 18-20			&	&	&		&	&	&		&	&n.s.	& \\
~~~~ 31-40			&	&	&		&	&	&		& 0.22	&***	&0.04 \\
~~~~ 41-50			&	&	&		&	&	&		& 0.31	&***	&0.05 \\
~~~~ 51 and older		&	&	&		&	&	&		& 0.39	&***	&0.05 \\
Is born abroad			&	&	&		&	&	&		&	&n.s.	& \\
Is female offender		&	&	&		&	&	&		&	&n.s.	& \\
Is repeat offender		&	&	&		&	&	&		& 0.11	&**	&0.03 \\
Has multiple victims		&	&	&		&	&	&		& 0.28	&***	&0.03 \\
Has offender basic skills	&	&	&		&	&	&		& 0.15	&***	&0.04 \\
Has offender special skills	&	&	&		&	&	&		& 0.67	&***	&0.10 \\
\noalign{\smallskip}\hline\noalign{\smallskip}
Adjusted $R^2$			& \multicolumn{3}{r}{0.31}	& \multicolumn{3}{r}{0.36}	& \multicolumn{3}{r}{0.39} \\
\multicolumn{10}{l}{Significance: * p $<$ 0.05, ** p $<$ 0.01, *** p $<$ 0.001, ~~~~~~ N=4674 (Valid listwise)} \\
\noalign{\smallskip}\hline
\end{tabular}
}
\end{table}

\subsection{Regression: High-tech crime leads to a significantly longer sentence}
\label{subsec:regression}
Before presenting detailed results of the analysis, we check whether the assumptions of the regression were met.

\begin{itemize}
\item There was independence of residuals, as assessed by a Durbin-Watson statistic of 1.75.

\item The independent variables are likely to be linearly related to the dependent variables as shown by a scatterplot of the studentized residuals against the unstandardised predicted values.

\item Visual inspection of the plot of the standardized residuals versus unstandardized predicted values revealed homoscedasticity.

\item There was no multicollinearity because the tolerance values were at least 0.34, which is well greater than 0.1.

\item There were 59 outliers with a sentence of, on average, 3.6 days whose residuals differed more from the predicted value than three standard deviations.
In 47 cases, a short prison sentence was imposed in addition to community service to comply with article 22b of the Penal code.
In 2 cases, the offender was partially excused by insanity defence and sentenced to a short prison sentence and a treatment.
In the remining 10 cases the judge imposed a mild sentence.
The outliers represent correctly coded judgments are were therefore kept in the dataset.
\end{itemize}

Table~\ref{tab:regression} shows that the Model 3 predicts the length of the custodial sentence in a statistically significant manner, F (27, 4646) = 111.85, p $<$0.0005, adjusted $R^2$ = 0.39.
Most independent variables contribute significantly to the result.

Table~\ref{tab:regression} shows that most of the variance in sentencing is explained by Model 1.
Models 2 and 3 each explain a small proportion of the variance.
This means that the sentencing decision depends mostly on the offense, as required by law.

Model 3 mainly confirms what we expect about the relationship between the length of the custodial sentence and the independent variables.
We give a few examples:

\begin{itemize}
% exp(B)=0.47	Maximum prison months 71 and shorter
% exp(B)=0.44	Maximum prison months 72-95
% exp(B)=0.73	Maximum prison months 96-107
% exp(B)=1.00	Maximum prison months 120-143
% exp(B)=1.55	Maximum prison months 144-179
% exp(B)=2.95	Maximum prison months 180-215
% exp(B)=3.01	Maximum prison months 216 and longer
\item The least serious category of offences leads, on average, to a 53\% shorter custodial sentence than an offence with the maximum sentence in the reference category (108-119 months).
\item The most serious category of offences leads on average to a 202\% longer custodial sentence than an offence with the maximum sentence in the reference category (108-119 months).

% exp(B)=0.89	Property offence
% exp(B)=0.46	Public order offence
% exp(B)=0.59	Other Penal code offence
% exp(B)=0.74	Road traffic offence
% exp(B)=0.47	Drugs related offence
% exp(B)=0.86	Weapons related offence
% exp(B)=0.75	Other criminal offence
% exp(B)=1.10	Number of offences
\item All offences lead to shorter sentences than an offence in the reference category (violent offenses).
\item For each additional offence being committed, the sentence length is increased, on average, by 10\%.

% exp(B)=1.11	Have guidelines been mentioned
% exp(B)=1.78	Has prosecution special expertise
\item If the court mentions guidelines in the judgment, the sentence is, on average, 11\% longer than a sentence where guidelines have not been consulted.
\item If the prosecution has engaged the services of specialists, the sentence is, on average, 78\% longer than sentences where specialists have not been involved.

% exp(B)=0.90	Age at sentencing 18-20
% exp(B)=1.24	Age at sentencing 31-40
% exp(B)=1.37	Age at sentencing 41-50
% exp(B)=1.48	Age at sentencing 51 and older
\item The youngest offenders, on average, receive a 10\% shorter sentence than offenders in the reference age group (21-30).
\item The oldest offenders, on average, receive a 48\% longer sentence than offenders in the reference age group (21-30).

% exp(B)=1.12	Is repeat offender
% exp(B)=1.33	Has multiple victims
\item Repeat offenders receive, on average, a 12\% longer sentence than first-time offenders.

% exp(B)=1.16	Has offender basic skills
% exp(B)=1.95	Has offender special skills
\item Offenders with special ICT skills receive, on average, a 95\% longer custodial sentence than offenders without special skills.
\end{itemize}

Factor analysis showed that the correlation between the manifest variables was always smaller than 0.25.
This makes it unlikely that there are latent factors that provide a better explanation for the variance in sentencing than the manifest variables.

We also investigated possible interaction effects.
There is a possible interaction between \textit{Maximum prison months} and \textit{Has prosecution special expertise} because for the most serious offenses the court may more often call in the expertise of specialists.
There is also a possible interaction between \textit{Type of offence} and \textit{Has offender special skills} because special skills may be more useful for some offences (e.g. fraud) than for others (e.g. robbery).
The change in $R^2$ due to these two interactions was found to be significant but small ($<$0.004).

Finally, there is also possible interaction between \textit{Age at sentencing} and \textit{Is repeat offender} because repeat offenders are usually older than first offenders.
The change in $R^2$ due to this last interaction was not significant.

\begin{table}
\caption{Reliability of the coding (N=275)}
\label{tab:reliability}
\begin{tabular}{l *{6}{r} }
\hline\noalign{\smallskip}
Variable			&Accuracy&Kappa	&TP	&FP	&FN	&TN \\
\noalign{\smallskip}\hline\noalign{\smallskip}
Year of birth			&0.99	&0.96	&197	& 0	& 4	&74 \\
Legal basis			&0.88	&0.67	&195	&31	& 2	&47 \\
Court decision			&0.99	&	&271	& 3	& 1	& 0 \\
\noalign{\smallskip}\hline\noalign{\smallskip}
\multicolumn{7}{l}{Reliability after improvements} \\
\noalign{\smallskip}\hline\noalign{\smallskip}
Year of birth			&1.00	&0.99	&200	&0	& 1	&74 \\
Legal basis			&0.95	&0.84	&214	&12	& 2	&47 \\
Court decision			&0.99	&	&273	& 1	& 1	& 0 \\
\noalign{\smallskip}\hline
\end{tabular}
\end{table}

\subsection{Cohen's Kappa: High reliability of the automatic encoding}
\label{subsec:kappa}
It is common for multiple coders to compare their results and adjust them to an agreed coding scheme.
In that case, the reliability of the coding before and after the agreement is also stated.
We have followed a similar process, with the agreement leading to some improvements to the coding software.

Table~\ref{tab:reliability} shows the reliability of the coding of the year of birth, the decision of the court, and the legal basis, both before and after making the improvements.
The manual coding corresponds well with the coding of all variables by the program.

In the technical literature, it is common to state the accuracy of the coding.
In the social science literature, Cohen's Kappa is preferred, because in this metric the accuracy is corrected for chance hits.
We give both metrics, as well as the number of true/false positives/negatives.
(In case TN = 0, Kappa cannot be calculated.)

% These variables have therefore not been checked:
% Have guidelines been mentioned,
% Has prosecution special expertise,
% Is born abroad,
% Is female offender,
% Is repeat offender,
% Has multiple victims,
% Has offender basic skills,
% Has offender special skills.

The main improvements made as a result of the reliability study are:

\begin{itemize}
\item The age of the offenders was sometimes stated in words, in which case the age was incorrectly missing in the coding (FN).

\item Several laws were missing, and there were some problems with article numbers so that the legal basis was sometimes incompletely coded (FP).

\item A rarely used type of court decision was missing and the decision was sometimes incomplete (FP).
\end{itemize}

\section{Case study: Discussion}
\label{sec:discussion}

\paragraph{Interpretation of the results} In related work, voluntary sentencing guidelines~\citep{Ulmer2004} and mandatory presumptive sentences~\citep{Engen2006} play an important role in harmonizing the sentence length.
These variables also explain a large part of the variance in sentence lengths.
In the Netherlands, sentencing guidelines are not universally applied.
This explains why they have a relatively small impact on sentence length in the models.

The average custodial sentence of the case study is about twice as long (24.8 months) as in the related work based on the Dutch data (see Table~\ref{tab:related_work}).
A possible, but unlikely explanation is that the Dutch courts have doubled the sentences in the last 10 years.
A more plausible explanation is that the offences present in the open data set are more serious cases than those studied by for example \citet{Wingerden2016}.
This is consistent with the criteria for publication as listed in Section~\ref{subsec:open_data} and the comparison from Table~\ref{tab:offences}.

Like \citet{Wingerden2016} we found that the youngest offenders receive shorter sentences while the oldest offenders receive longer sentences.
This can be explained by observing that an older offender is more likely to be a repeat offender and that a younger offender is usually treated more leniently.

The maximum sentence length is not the best predictor of the sentence length.
For example, we have found that the least serious offences lead to shorter sentences, and the most serious offences lead to longer sentences.
However, there is often a big gap between the maximum sentence and the actual sentence length.
To illustrate this, we give a few examples:
\begin{itemize}
\item Human trafficking (273f Penal code) has a maximum of 30 years, but an employee of a swingers club was sentenced to a prison sentence of 1 day for neglecting to verify the age of a minor visitor (ECLI:NL:RBOVE:2015:5781).

\item Assault (302 Penal code) carries a maximum of 10 years, but a student was sentenced to 31 days in prison for assault during hazing (ECLI:NL:RBNNE\discretionary{:}{}{:}2017:4461).

\item Possession of certain illegal narcotics (articles 2 and 10 Opium Act) carries a maximum sentence of 6 years imprisonment, but an offender who threw 4.2 grams of MDMA over the walls of a prison was sentenced to 1 week detention (ECLI:NL:RBROT:2017:6082).
\end{itemize}
A large gap between maximum sentence and actual sentence creates outliers in the regression, which reduces the power of the analysis.

In contrast to the results of \citet{Wingerden2016}, the gender of the offender is not significant.
There are two possible explanations for this.
First, our coding program may make mistakes in determining the sex of the offender.
We tried to rule out this contingency by manually checking a randomized, stratified sample.
Second, the percentage of women in the case study (8\%) is even lower than in \citet{Wingerden2016} (10\%).
Women are underrepresented in serious crimes~\citep{Daly1997}, and there may be too few women in our data set to achieve a statistically significant result.

\paragraph{Contribution to the theory}
The results show that offenders in a high-tech case receive a more severe sentence than offenders in a low-tech case.
An explanation could be that offenders with special skills find it realitively easy to commit large-scale offences; the offender uses ICT as a ``crime multiplier''.
Therefore we searched the dataset for court judgments with the word \textit{large-scale} (N=1517).
The word occurs in 30.8\% of the court judgments against an offender with special skills,
and 5.8\% of the court judgments against an offender without special skills.
This difference in frequencies is statistically significant ($\chi^2(1)=1039.0, p<0.0005$).
We give three examples of offences where ICT is used as a crime multiplier.
\begin{description}
\item[Drugs] The street dealer must devote considerable time and effort to maintaining a relationship with his customers without attracting the attention of the police.
By contrast, the vendor on a darknet marketplace does not have to spend time and effort to avoid rasing suspicion because the technology (the TOR network and Bitcoin) does that for him.
The technology not only makes it difficult for police to get a grip on the vendor, but it also makes a large-scale and global approach to drug trade easy~\citep{Christin2013}.

\item[CSA] Not long ago, offenders of child sexual abuse had to rely on physical media.
With the advent of the Internet, it became possible to collect and share media on a large scale, whereby the technology ensures anonymity and scalability~\citep{Bruggen2021}.

\item[Fraud] Online banking is not only cost-effective for banks but also fraudsters.
With a large-scale and inexpensive phishing campaign, an offender can reach millions of potential victims.
In this scenario, the offender only uses technology to achieve scalability.
To avoid detection, he uses money mules~\cite{Leukfeldt2014}.
\end{description}

In each of these examples, offenders use ICT as a crime multiplier to increase the scale of the offence.
This not only makes offenders with special skills more blameworthy (because they make more victims), it also makes offenders with special skills more dangerous (because if they re-offend, they have the skills to do so on a large scale).
The explanation, therefore, of the main result of the case study is that special skills is a factor for two out of three focal concerns.
Future research is needed to validate this refinement of focal concerns theory.

\paragraph{Limitations}
The published court judgments from the Open data of \href{https://www.rechtspraak.nl/Uitspraken}{rechtspraak.nl} relate to serious crimes.
Therefore, our results are not representative of the total population of criminal court judgments.

The automatic coding of court judgments can lead to incorrect or missing coding.
We manually checked a stratified random sample to estimate the error margins, which are small but not zero.

The public data that we have analysed contains little personal information due to the European privacy laws.
Related work generally uses richer data, so that the $R^2$ is generally higher in linear regression models that predict the length of the custodial sentence.
For example, (1) information about the education, marital status, and family composition of the offender~\citep{Blowers2013},
(2) the number of previous convictions, and country of origin~\citep{Wingerden2016},
and (3) the education and social-economic status~\citep{Spohn2014} of the offender all help to explain a greater proportion of variance in sentencing.
Such information is not available in the court judgments of \href{https://www.rechtspraak.nl/Uitspraken}{rechtspraak.nl}.

% \cbstart
Most researchers have full control over the selection of court judgments they want to analyse.
Because only a fraction of the court judgments are available as open data, we had less control over the selection.
This makes it difficult to reach a conclusion about the representativeness of the results.
% \cbend

\section{Conclusions}
\label{sec:conclusions}
Transparency promotes confidence in the judiciary.
Because open data promotes transparency, some countries make judicial documentation publicly available as open data.
In the Netherlands, about 5\% of the criminal court judgments were published for several years.
Due to European privacy legislation, personal information is removed from the court judgments before being made public.

One of the goals of research in sentencing severity is to contribute to the transparency of the judiciary.
Since open data and sentencing research are both aimed at transparency, it is interesting to investigate sentencing severity using open data.
However, this poses a challenge that we are addressing in the paper.
It is normal for the personal circumstances of the offender to be taken into account by the judge when making the decision.
Repeat offenders, for example, often receive a longer prison sentence than first-time offenders, because repeat offenders pose a greater danger to society.
But if open data does not contain personal information, is it still possible to conduct sentencing research on open data? 
Through sentencing research, it can be established to what extent the judges deviate from the norms.
This also promotes transparency.

\paragraph{Research questions} We answer the first question ``To what extent are the intrinsic limitations of open data a barrier to obtaining good results?'' by performing a case study.
The open data comes from the Dutch criminal justice system~\href{https://www.rechtspraak.nl/Uitspraken}{Rechtspraak.nl}.
We analysed 25,366 court judgments from the period 2015-2020 and in particular, examined the relationship between the severity of the sentence and a range of predictors.
The main result is that the quality of sentencing research with open data is comparable to the quality of sentencing research on data composed of files that have not been anonymised.
% \cbstart
The answer to the research question is thus a qualified yes, in the sense that the variance explained by the linear regression model of the case study is 39\%, whereas comparable related work reports 60\%~\citep{Wingerden2016}.
We consider this a good result because we had no control over the selection of the court judgments, and because we had strictly less information per case.
% \cbend

The second research question ``To what extent do offenders in high-tech cases receive a more severe sentence than offenders in low-tech cases?'' has been investigated in the case study.
We found that offenders who use advanced ICT are sentenced to significantly longer custodial sentences compared to other offenders.
The explanation is that ICT can be used as a crime multiplier, which makes an offender in high-tech cases both more blameworthy and more dangerous.
We propose to include technical skills as one of the factors that determine blameworthiness and dangerousness in focal concerns theory.

\paragraph{Policy implication} We provide two policy recommendations for the judiciary to make open data even more valuable.
Firstly, court judgments are structured, but this structure is not strictly enforced~\citep{Trompper2016}.
Natural language offers too many possibilities to vary even standard structures so that building a coding program is still time-consuming.
It would be advisable for \href{https://www.rechtspraak.nl/Uitspraken}{rechtspraak.nl} to enrich the meta-data already present in the court judgments with substantive information about the offender, the offenses, the legal basis, and the decision.
In Appendix~\ref{app:coding}, we provide a concrete proposal for enriched meta-data.

Secondly, we recommend including in the meta-data also the specific publication criteria that were applicable when the court judgment was published.
This would make it possible to calculate more precisely to what extent the open data from \href{https://www.rechtspraak.nl/Uitspraken}{rechtspraak.nl} is representative for all criminal court judgments in the Netherlands.

We found several errors in published court judgments and reported them to the appropriate district courts.

% \begin{acknowledgements}
\section*{Acknowledgments}
We thank Marianne Junger and Mortaza Shoae Barg 
for their help in writing this paper. \\[1ex]
\textbf{Conflict of interest} The authors have no conflict of interest with regard to the research, authorship, and/or publication of this article. \\[1ex]
\textbf{Funding} The authors have not received financial support for the research, authorship and/or publication of this article. \\[1ex]
\textbf{Availability of data} The full data set is available as open data from \href{https://www.rechtspraak.nl/Uitspraken}{rechtspraak.nl}. \\[1ex]
% \cbstart
\textbf{Code availability} The coding software that we developed for this paper is available from\\
\href{https://doi.org/10.5281/zenodo.4666679}{https://doi.org/10.5281/zenodo.4666679} \\[1ex]
\textbf{Compliance with ethical standards} The research complies with ethical standards because all data that has been analysed is open data that has been made public for analysis by third parties.
% \cbend
% \end{acknowledgements}

\bibliographystyle{apalike}
\bibliography{darkweb_refs}

\begin{thebibliography}{}

\bibitem[Abrams, 2011]{Abrams2011}
Abrams, D.~S. (2011).
\newblock Is pleading really a bargain?
\newblock {\em J. of Empirical Legal Studies}, 8(S1):200--221.
\newblock DOI:10.1007/S12103-012-9167-1.

\bibitem[Albonetti, 1997]{Albonetti1997}
Albonetti, C.~A. (1997).
\newblock Sentencing under the federal sentencing guidelines: Effects of
  defendant characteristics, guilty pleas, and departures on sentence outcomes
  for drug offenses, 1991-1992.
\newblock {\em Law \& Society Review}, 31(4):789--822.
\newblock DOI:10.2307/3053987.

\bibitem[Bargh et~al., 2017]{Bargh2017}
Bargh, M.~S., Choenni, S., and Meijer, R.~F. (2017).
\newblock Integrating semi-open data in a criminal judicial setting.
\newblock In {\em Achieving Open Justice through Citizen Participation and
  Transparency}, pages 137--156. IGI Global.
\newblock DOI:10.4018/978-1-5225-0717-8.ch007.

\bibitem[Berdejs\'o and Yuchtman, 2013]{Berdejo2013}
Berdejs\'o, C. and Yuchtman, N. (2013).
\newblock Crime, punishment, and politics: an analysis of political cycles in
  criminal sentencing.
\newblock {\em Review of Economics and Statistics}, 95(3):741--756.
\newblock DOI:10.1162/REST\_a\_00296.

\bibitem[Bijlenga and Kleemans, 2018]{Bijlenga2018}
Bijlenga, N. and Kleemans, E.~R. (2018).
\newblock Criminals seeking {ICT}-expertise: an exploratory study of {Dutch}
  cases.
\newblock {\em European J. on Criminal Policy and Research}, 24:253--268.
\newblock DOI:10.1007/s10610-017-9356-z.

\bibitem[Blowers and Doerner, 2015]{Blowers2013}
Blowers, A.~N. and Doerner, J.~K. (2015).
\newblock Sentencing outcomes of the older prison population: an exploration of
  the age leniency argument.
\newblock {\em J. of Crime and Justice}, 38(1):58--76.
\newblock DOI:10.1080/0735648X.2013.822161.

\bibitem[Cao et~al., 2020]{Cao2020}
Cao, Y., Ash, E., and Chen, D.~L. (2020).
\newblock Automated fact-value distinction in court opinions.
\newblock {\em European Journal of Law and Economics}, 50:451--467.
\newblock DOI:10.1007/s10657-020-09645-7.

\bibitem[Christin, 2013]{Christin2013}
Christin, N. (2013).
\newblock Traveling the {Silk Road}: A measurement analysis of a large
  anonymous online marketplace.
\newblock In {\em Int. Conf. on World Wide Web (WWW)}, pages 213--224, Rio de
  Janeiro, Brazil. ACM, New York.
\newblock DOI:10.1145/2488388.2488408.

\bibitem[Cohen and Yang, 2019]{Cohen2019}
Cohen, A. and Yang, C.~S. (2019).
\newblock Judicial politics and sentencing decisions.
\newblock {\em American Economic J.: Economic Policy}, 11(1):160--191.
\newblock DOI:10.1257/pol.20170329.

\bibitem[Connolly and Wall, 2019]{Connolly2019}
Connolly, L.~Y. and Wall, D.~S. (2019).
\newblock The rise of crypto-ransomware in a changing cybercrime landscape:
  Taxonomising countermeasures.
\newblock {\em Computers \& Security}, 87:1--18.
\newblock DOI:10.1016/j.cose.2019.101568.

\bibitem[Daly and Tonry, 1997]{Daly1997}
Daly, K. and Tonry, M. (1997).
\newblock Gender, race, and sentencing.
\newblock {\em Crime and Justice}, 22:201--252.
\newblock DOI:10.1086/449263.

\bibitem[Doherty and Steinberg, 2016]{Doherty2016}
Doherty, J.~W. and Steinberg, R.~H. (2016).
\newblock Punishment and policy in international criminal sentencing: an
  empirical study.
\newblock {\em American J. of Int. Law}, 110(1):49--81.
\newblock DOI:10.5305/amerjintelaw.110.1.0049.

\bibitem[Engen and Gainey, 2000]{Engen2006}
Engen, R.~L. and Gainey, R.~R. (2000).
\newblock Modeling the effects of legally relevant and extralegal factors under
  sentencing guidelines: The rules have changed.
\newblock {\em Criminology}, 38(4):1207--1230.
\newblock DOI:10.1111/j.1745-9125.2000.tb01419.x.

\bibitem[Feldmeyer and Ulmer, 2011]{Feldmeyer2011}
Feldmeyer, B. and Ulmer, J.~T. (2011).
\newblock Racial/ethnic threat and federal sentencing.
\newblock {\em J. of Research in Crime and Delinquency}, 48(2):238--270.
\newblock DOI:10.1177/0022427810391538.

\bibitem[Freiburger and Hilinski, 2010]{Freiburger2010}
Freiburger, T.~L. and Hilinski, C.~M. (2010).
\newblock The impact of race, gender, and age on the pretrial decision.
\newblock {\em Criminal Justice Review}, 35(3):318--334.
\newblock DOI:10.1177/0734016809360332.

\bibitem[Gottschalk, 2014]{Gottschalk2014}
Gottschalk, P. (2014).
\newblock Crime: The amount and disparity of sentencing - a comparison of
  corporate and occupational white collar criminals.
\newblock {\em Int. J. of Law, Crime and Justice}, 42(3):175--187.
\newblock DOI:10.1016/j.ijlcj.2014.01.002.

\bibitem[Hadzhidimova and Payne, 2019]{Hadzhidimova2019}
Hadzhidimova, L.~I. and Payne, B.~K. (2019).
\newblock The profile of the international cyber offender in the {U.S.}
\newblock {\em Int. J. of Cybersecurity Intelligence \& Cybercrime},
  2(1):40--55.
\newblock https://vc.bridgew.edu/ijcic/vol2/iss1/4.

\bibitem[Hartley, 2014]{Hartley2014}
Hartley, R.~D. (2014).
\newblock Focal concerns theory.
\newblock In {\em The Encyclopedia of Theoretical Criminology}, pages 1--5.
  Blackwell Publishing.
\newblock DOI:10.1002/9781118517390.wbetc148.

\bibitem[Hartley et~al., 2011]{Hartley2011a}
Hartley, R.~D., Kwak, D.-H., Park, M., and Lee, M.-S. (2011).
\newblock Exploring sex disparity in sentencing outcomes: A focus on narcotics
  offenders in {South Korea}.
\newblock {\em Int. J. of Offender Therapy and Comparative Criminology},
  55(2):268--286.
\newblock DOI:10.1177/0306624X09360966.

\bibitem[Hester and Hartman, 2017]{Hester2017}
Hester, R. and Hartman, T.~K. (2017).
\newblock Conditional race disparities in criminal sentencing: A test of the
  liberation hypothesis from a non-guidelines state.
\newblock {\em J. of quantitative criminology}, 33:77--100.
\newblock DOI:10.1007/s10940-016-9283-z.

\bibitem[Holtfreter, 2013]{Holtfreter2013}
Holtfreter, K. (2013).
\newblock Gender and "other people's money": An analysis of white-collar
  offender sentencing.
\newblock {\em Women \& Criminal Justice}, 23(4):326--344.
\newblock DOI:10.1080/08974454.2013.821015.

\bibitem[Huang et~al., 2010]{Huang2010}
Huang, K.-C., Chen, K.-P., and Lin, C.-C. (2010).
\newblock Does the type of criminal defense counsel affect case outcomes?: A
  natural experiment in {Taiwan}.
\newblock {\em Int. Review of Law and Economics}, 30(2):113--127.
\newblock DOI:10.1016/j.irle.2009.09.005.

\bibitem[Kahneman, 2011]{Kahneman2011}
Kahneman, D. (2011).
\newblock {\em Thinking, Fast and Slow}.
\newblock Penguin.

\bibitem[Koons-Witt et~al., 2014]{Koons-Witt2014}
Koons-Witt, B.~A., Sevigny, E.~L., Burrow, J.~D., and Hester, R. (2014).
\newblock Gender and sentencing outcomes in {South Carolina}: Examining the
  interactions with race, age, and offense type.
\newblock {\em Criminal Justice Policy Review}, 25(3):299--324.
\newblock DOI:10.1177/0887403412468884.

\bibitem[Lee et~al., 2011]{Lee2011}
Lee, M., Ulmer, J.~T., and Park, M. (2011).
\newblock Drug sentencing in {South Korea}: The influence of case-processing
  and social status factors in an ethnically homogeneous context.
\newblock {\em J. of Contemporary Criminal Justice}, 27(3):378--397.
\newblock DOI:10.1177/1043986211412574.

\bibitem[Leifker and Sample, 2011]{Leifker2011}
Leifker, D. and Sample, L.~L. (2011).
\newblock Probation recommendations and sentences received: The association
  between the two and the factors that affect recommendations.
\newblock {\em Criminal Justice Policy Review}, 22(4):494--517.
\newblock DOI:10.1177/0887403411388405.

\bibitem[Leukfeldt, 2014]{Leukfeldt2014}
Leukfeldt, E.~R. (2014).
\newblock Cybercrime and social ties -- phishing in amsterdam.
\newblock {\em Trends in Organized Crime}, 17(4):231--249.
\newblock DOI:10.1007/s12117-014-9229-5.

\bibitem[Light and Wermink, 2021]{Light2021}
Light, M.~T. and Wermink, H. (2021).
\newblock The criminal case processing of foreign nationals in the
  {Netherlands}.
\newblock {\em European Sociological Review}, 37(1):104--120.
\newblock DOI:10.1093/esr/jcaa039.

\bibitem[LOVS, 2020]{LOVS2020}
LOVS (2020).
\newblock Ori\"entatiepunten voor straftoemeting en lovs-afspraken.
\newblock Technical report, Landelijk Overleg Vakinhoud Strafrecht.
\newblock
  https://www.rechtspraak.nl/SiteCollectionDocuments/Orientatiepunten-en-afspraken-LOVS.pdf.

\bibitem[Lynch and Omori, 2018]{Lynch2018}
Lynch, M. and Omori, M. (2018).
\newblock Crack as proxy: Aggressive federal drug prosecutions and the
  production of black-white racial inequality.
\newblock {\em Law \& Society Review}, 52(3):773--809.
\newblock DOI:10.1111/lasr.12348.

\bibitem[Maddan et~al., 2012]{Maddan2012}
Maddan, S., Hartley, R.~D., Walker, J.~T., and Miller, J.~M. (2012).
\newblock Sympathy for the devil: An exploration of federal judicial discretion
  in the processing of white-collar offenders.
\newblock {\em American J. of Criminal Justice}, 37:4--18.
\newblock DOI:10.1007/s12103-010-9094-y.

\bibitem[Marcum et~al., 2011]{Marcum2011}
Marcum, C.~D., Higgins, G.~E., and Tewksbury, R. (2011).
\newblock Doing time for cyber crime: An examination of the correlates of
  sentence length in the {United States}.
\newblock {\em Int. J. of Cyber Criminology}, 5(2):824--835.
\newblock https://www.cybercrimejournal.com/marcumetal2011julyijcc.pdf.

\bibitem[McAdams, 2016]{McAdams2016}
McAdams, J.~M. (2016).
\newblock The effect of school starting age policy on crime: Evidence from {US}
  microdata.
\newblock {\em Economics of Education Review}, 54:227--241.
\newblock DOI:10.1016/j.econedurev.2014.12.001.

\bibitem[McEwen and Regoeczi, 2015]{McEwen2015}
McEwen, T. and Regoeczi, W. (2015).
\newblock Forensic evidence in homicide investigations and prosecutions.
\newblock {\em J. of forensic sciences}, 60(5):1188--1198.
\newblock DOI:10.1111/1556-4029.12787.

\bibitem[Meijer et~al., 2020]{Meijer2020}
Meijer, R.~F., van~den Braak, S.~W., and Choenni, S. (2020).
\newblock Criminaliteit en rechtshandhaving 2019 ontwikkelingen en samenhangen.
\newblock Cahier 2020-16, Wetenschappelijk Onderzoek- en Documentatiecentrum
  (WODC).
\newblock https://repository.wodc.nl/handle/20.500.12832/254.

\bibitem[Montoya et~al., 2013]{Montoya2013}
Montoya, L., Junger, M., and Hartel, P. (2013).
\newblock How 'digital' is traditional crime?
\newblock In {\em European Intelligence and Security Informatics Conference},
  pages 31--37, Uppsala, Sweden. IEEE.
\newblock DOI:10.1109/EISIC.2013.12.

\bibitem[Patrick and Marsh, 2011]{Patrick2011}
Patrick, S. and Marsh, R. (2011).
\newblock Sentencing outcomes of convicted child sex offenders.
\newblock {\em J. of child sexual abuse}, 20(1):94--108.
\newblock DOI:10.1080/10538712.2011.541356.

\bibitem[Peterson et~al., 2013]{Peterson2013}
Peterson, J.~L., Strom, K.~J., and Johnson, D.~J. (2013).
\newblock Effect of forensic evidence on criminal justice case processing.
\newblock {\em J. of forensic science}, 58(S1):S78--S90.
\newblock DOI:10.1111/1556-4029.12020.

\bibitem[Pina-S\'anchez et~al., 2019]{Pina-Sanchez2019a}
Pina-S\'anchez, J., Grech, D., Brunton-Smith, I., and Sferopoulos, D. (2019).
\newblock Exploring the origin of sentencing disparities in the crown court:
  Using text mining techniques to differentiate between court and judge
  disparities.
\newblock {\em Social Science Research}, 84:102371:1--13.
\newblock DOI:10.1016/j.ssresearch.2019.102343.

\bibitem[Pina-S\'anchez and Linacre, 2014]{Pina-Sanchez2014}
Pina-S\'anchez, J. and Linacre, R. (2014).
\newblock Enhancing consistency in sentencing: Exploring the effects of
  guidelines in {England} and {Wales}.
\newblock {\em J. of Quantitative Criminology}, 30:731--748.
\newblock DOI:10.1007/s10940-014-9221-x.

\bibitem[Raets and Janssens, 2019]{Raets2019}
Raets, S. and Janssens, J. (2019).
\newblock Trafficking and technology: Exploring the role of digital
  communication technologies in the {Belgian} human trafficking business.
\newblock {\em European J. on Criminal Policy and Research}, Online first.
\newblock DOI:10.1007/s10610-019-09429-z.

\bibitem[Rydberg et~al., 2018]{Rydberg2018}
Rydberg, J., Cassidy, M., and Socia, K.~M. (2018).
\newblock Punishing the wicked: Examining the correlates of sentence severity
  for convicted sex offenders.
\newblock {\em J. of quantitative criminology}, 34:943--970.
\newblock DOI:10.1007/s10940-017-9360-y.

\bibitem[Simon, 1955]{Simon1955}
Simon, H.~A. (1955).
\newblock A behavioral model of rational choice.
\newblock {\em The Quarterly J. of Economics}, 69(1):99--118.
\newblock http://www.jstor.org/stable/1884852.

\bibitem[Snodgrass et~al., 2011]{Snodgrass2011}
Snodgrass, G.~M., Blokland, A. A.~J., Haviland, A., Nieuwbeerta, P., and Nagin,
  D.~S. (2011).
\newblock Does the time cause the crime? an examination of the relationship
  between time served and reoffending in the {Netherlands}.
\newblock {\em Criminology}, 49(4):1149--1194.
\newblock DOI:10.1111/j.1745-9125.2011.00254.x.

\bibitem[Spohn et~al., 2014]{Spohn2014}
Spohn, C.~C., Kim, B., Belenko, S., and Brennan, P.~K. (2014).
\newblock The direct and indirect effects of offender drug use on federal
  sentencing outcomes.
\newblock {\em J. of quantitative Criminology}, 30:549--576.
\newblock DOI:10.1007/s10940-014-9214-9.

\bibitem[Trompper and Winkels, 2016]{Trompper2016}
Trompper, M. and Winkels, R. (2016).
\newblock Automatic assignment of section structure to texts of {Dutch} court
  judgments.
\newblock In {\em Legal Knowledge and Information Systems: JURIX 2016: The
  Twenty- Ninth Annual Conference}, volume Frontiers in Artificial Intelligence
  and Applications 294, pages 167--172, Sophia Antipolis, France. IOS Press,
  Amsterdam.
\newblock DOI:10.3233/978-1-61499-726-9-167.

\bibitem[Ulmer and Johnson, 2004]{Ulmer2004}
Ulmer, J.~T. and Johnson, B. (2004).
\newblock Sentencing in context: A multilevel analysis.
\newblock {\em Criminology}, 42(1):137--178.
\newblock DOI:10.1111/j.1745-9125.2004.tb00516.x.

\bibitem[van~den Hoven, 2011]{Hoven2011}
van~den Hoven, P. (2011).
\newblock Een strafrechtelijke uitspraak als tekstueel stelsel.
\newblock {\em Tijdschrift voor Taalbeheersing}, 33(1):5--15.
\newblock DOI:10.5117/TVT2011.1.EEN\_397.

\bibitem[van~der Bruggen and Blokland, 2021]{Bruggen2021}
van~der Bruggen, M. and Blokland, A. (2021).
\newblock A crime script analysis of child sexual exploitation material fora on
  the darkweb.
\newblock {\em Sexual abuse}, Online first.
\newblock DOI:10.1177/1079063220981063.

\bibitem[van Wingerden et~al., 2016]{Wingerden2016}
van Wingerden, S., van Wilsem, J., and Johnson, B.~D. (2016).
\newblock Offender's personal circumstances and punishment: Toward a more
  refined model for the explanation of sentencing disparities.
\newblock {\em Justice Quarterly}, 33(1):100--133.
\newblock DOI:10.1080/07418825.2014.902091.

\bibitem[Viglione et~al., 2011]{Viglione2011}
Viglione, J., Hannon, L., and DeFina, R. (2011).
\newblock The impact of light skin on prison time for black female offenders.
\newblock {\em The Social Science J.}, 48(1):250--258.
\newblock DOI:10.1016/j.soscij.2010.08.003.

\bibitem[Wermink et~al., 2017]{Wermink2017}
Wermink, H., Johnson, B.~D., de~Keijser, J.~W., Dirkzwager, A. J.~E., Reef, J.,
  and Nieuwbeerta, P. (2017).
\newblock The influence of detailed offender characteristics on consecutive
  criminal processing decisions in the {Netherlands}.
\newblock {\em Crime \& Delinquency}, 63(10):1279--1313.
\newblock DOI:10.1177/0011128715624929.

\bibitem[Wermink et~al., 2015]{Wermink2015}
Wermink, H., Johnson, B.~D., Nieuwbeerta, P., and de~Keijser, J.~W. (2015).
\newblock Expanding the scope of sentencing research: Determinants of juvenile
  and adult punishment in the {Netherlands}.
\newblock {\em European Journal of Criminology}, 12(6):739--768.
\newblock DOI:10.1177/1477370815597253.

\bibitem[Wu and DeLone, 2012]{Wu2012}
Wu, J. and DeLone, M.~A. (2012).
\newblock Revisiting the normal crime and liberation hypotheses: Citizenship
  status and unwarranted disparity.
\newblock {\em Criminal Justice Review}, 37(2):214--238.
\newblock DOI:10.1177/0734016811436336.

\end{thebibliography}

\appendix

\section{Appendix: Automatic coding}
\label{app:coding}
We have written a JavaScript program that can automatically code criminal court judgments downloaded as an XML file from \href{https://www.rechtspraak.nl/Uitspraken}{rechtspraak.nl}.
The program performs the following steps for each XML file:

\paragraph{Read input}
\label{subsec:input}
Read an XML file and convert the court judgment to plain text by removing the diacritics, redundant spacing, etc.

\paragraph{Layout}
\label{subsec:layout}
Determine the chapter structure of the court judgment.
Usually, a judgment contains \textit{$<$section$>$} XML tags that identify the standard chapters, such as \textit{Wettelijke voorschriften}, and \textit{Beslissing}, but not always.
If the \textit{$<$section$>$} tags provide insufficient information, the program will search other XML tags, such as \textit{$<$title$>$}, \textit{$<$bridgehead$>$}, and \textit{$<$emphasis$>$} for chapter titles.
The program takes into account different ways of spelling the chapter titles.
For example, \textit{De op te leggen maatregel berust op artikel}, \textit{De Wet}, and \textit{Wettelijke voorschriften} all refer to the same chapter.

\paragraph{Extra-legal information}
\label{subsec:extra_legal}
Search all chapters for extra-legal information related to the offender, such as year of birth, country of birth, and sex.
In many court judgments the year of birth is missing, but often the press release that is also part of the XML file does state the age of the offender.
The sex of the offender is never explicitly stated, but from phrases such as \textit{verdachte en haar medeverdachte} the program deduces that offender is a woman.
Without such indications, the program assumes that the offender is a man.

\paragraph{Legal information}
\label{subsec:legal}
Search all chapters for legal variables, such as first/repeat offender, number of co-offenders, victims, etc.
There are several ways to indicate whether the offender is a first or repeat offender.
For example, \textit{in het voordeel van de verdachte is gelet op zijn blanco strafblad}, \textit{verdachte is first offender}, indicate that offender is first offender.
A fragment such as \textit{volgens justitiele documentatie heeft verdachte zich eerder schuldig gemaakt} indicates that the offender is a repeat offender.

\paragraph{Technical information regarding the offender}
\label{subsec:offender}
The program searches the court judgments for a total of 20 different standard ICT concepts.
For example to code that a social network has played a role, the program searches for words such as as \textit{Facebook}, \textit{Instagram}, and \textit{LinkedIn}.
The program also looks for 19 different advanced ICT concepts.
For example, passages such as \textit{the onion router},\textit{TOR}, and \textit{hidden server} indicate that the offender has used the TOR network.
The program also searches the chapters for the occurrence of the most prominent marketplaces on the dark web, such as \textit{Silkroad}, \textit{Hansa}, and \textit{Evolution}.

\paragraph{Technical information regarding the prosecution}
\label{subsec:prosecution}
Search all chapters for indications of the technical expertise of the police and judicial authorities.
First, the chapters are searched to see if experts were involved in the case from institutions such as \textit{NFI}, \textit{TMFI}, and \textit{TNO}.
Secondly, it is checked whether seized equipment has been investigated by searching for \textit{gekraakt}, \textit{digitaal onderzoek}, or \textit{uitlezen}.
Third, the program determines whether a telco has been asked to provide data from a cell tower by searching for words such as \textit{aanstralen}.
Finally, the program checks whether there has been a telephone or internet tap by searching for text fragments such as \textit{artikel 126 Wetboek Strafvordering}, \textit{opnemen van vertrouwelijke communicatie}, en \textit{getapt door de politie}.

To know whether the court has consulted the sentencing guidelines, we search for \textit{LOVS}, \textit{orientatiepunt}, and the numbers and titles of the guidelines of the Public Prosecution Service.

\paragraph{Legal basis}
\label{subsec:legal_basis}
Look in the chapter \textit{Wettelijke voorschriften} for the articles of law on which the decision is based.
The laws and articles can appear in different orders, and they are also abbreviated in different ways.
For example: \textit{Sr art. 33, 33a, 47, 57, 420ter}, and \textit{27a Sr}.

\paragraph{The decision}
\label{subsec:decision}
In the chapter \textit{Beslissing}, look for the first decision mentioned.
There are several possibilities: measures, prison, community service, fine, acquittal, and procedural outcomes.
These can be issued in combination, and they can also be imposed partly conditionally.
We give one or more examples of ways in which the decision is recorded per type of decision:

\begin{itemize}
\item Measure:
\textit{Gelast de plaatsing van de verdachte in een psychiatrisch ziekenhuis voor een termijn van 1 (een) jaar},
\textit{Legt de verdachte op de maatregel tot plaatsing in een inrichting voor stelselmatige daders voor de duur van 2 jaren}, and
\textit{Maatregel - gelast ten aanzien van feit 1 subsidiair de terbeschikkingstelling van verdachte}.

\item Incarceration:
\textit{Beslissing: gevangenisstraf voor de duur van 159 dagen},
\textit{LEGT OP een gevangenisstraf voor de duur van 5 (vijf) jaren}, and
\textit{Veroordeelt verdachte tot een levenslange gevangenisstraf}.

\item Community service:
\textit{gelast het verrichten van een taakstraf voor de duur van 180 uren},
\textit{legt aan verdachte op de leerstraf So Cool voor de duur van 40 uren}, and
\textit{Veroordeelt verdachte tot het verrichten van 240 (tweehonderd veertig) uren taakstraf}.

\item Fine:
\textit{bepaalt de verplichting tot betaling aan de Staat ter ontneming van het wederrechtelijk verkregen voordeel op euro 5.700,-},
\textit{legt aan verdachte op een geldboete ten bedrage van 500 (vijfhonderd) euro}, and
\textit{veroordeelt de verdachte voor de bewezen verklaarde feiten tot een geldboete van euro 25.000,-}.

\item Acquittal:
\textit{Bepaalt dat ter zake van het bewezen verklaarde geen straf of maatregel wordt opgelegd},
\textit{niet strafbaar voor het bewezen verklaarde en ontslaat verdachte van alle rechtsvervolging}, and
\textit{Verklaart de ten laste gelegde feiten niet bewezen en spreekt verdachte daarvan vrij}.

\item Procedural:
\textit{Verklaart TOELAATBAAR de door Zwitsersland verzochte uitlevering van [opgeeiste persoon]},
\textit{gelast de tenuitvoerlegging van de voorwaardelijk opgelegde maatregel van plaatsing in een inrichting}, and
\textit{heropent de behandeling; - bepaalt dat de rechtbank de zaak als raadkamer voortzet; - schorst de behandeling}.
\end{itemize}

\paragraph{Analysis of the data in SPSS}
\label{subsec:analysis}
When the coding is done, the program produces a file with JSON records as output.
An example of the JSON record for a specific court judgment is shown below.
The JSON records of all court judgments are then converted into a CSV file that can be read into SPSS and analysed.
The example includes standard fields from the XML file, such as the ECLI identifier and the press release (\textit{Inhoudsindicatie}).
The fields that were added by the coding program are \textit{Geboortejaar} \dots \textit{Beslissing}.

\begin{verbatim}
{
    "ECLI": "ECLI:NL:RBMNE:2014:4790",
    "Datum_uitspraak": "09-10-2014",
    "Instelling": "Rechtbank Midden-Nederland",
    "Zaaknummer": "16/659159-14 (P)",
    "Type": "Uitspraak",
    "Locatie": "Utrecht",
    "Rechtsgebieden": ["Strafrecht"],
    "Taal": "nl",
    "Inhoudsindicatie": "De rechtbank Midden-Nederland veroordeelt
    	vijf verdachten tot gevangenisstraffen voor onder meer het
    	faciliteren en bevorderen van drugshandel via online marktplaatsen.
    	Via het beveiligde [naam]-netwerk maakten verdachten het mogelijk
    	om op de website [naam], en later op de website [naam], op anonieme
    	wijze drugs te kopen en te verkopen, ook buiten Nederland. De
    	rechtbank rekent het de verdachten zwaar aan dat zij gedurende
    	een langere periode bewust gedragingen hebben verricht die de
    	samenleving zeer ondermijnen. Drie verdachten waren actief als
    	moderators op [naam]. Door hun aandeel in de website hebben ze
    	de handel van verdovende middelen bevorderd. Deze verdachten
    	verkochten zelf ook drugs via de site. Verdachte en een 30-jarige
    	Enschedeer hebben naast hun werkzaamheden als moderator voor
    	[naam], ook de marktplaats [naam] laten ontwikkelen. De rechtbank
    	veroordeelt verdachte tot een gevangenisstraf van 6 jaar",
    "Geboortejaar": 1992,
    "Geboorteland": "buitenland",
    "Geslacht": "man",
    "Onderzoek": ["Commodore"],
    "Expertise_verdachte": ["Crypto_currency","Market","Wallet"],
    "Internet": ["Communication","Company","Generic","Hardware",
        "Mail","Mobile","Phone"],
    "Expertise_rechtbank": ["NFI"],
    "Opsporing": ["Infiltratie","Pseudo-koop","Tap-opname","Aanhouding"],
    "Verdachten_aantal": 5,
    "Recidive": "Eerste keer",
    "Wettelijke_voorschriften": [["Sr","33","33a","47","57"],
        ["Opw","10","10a"]],
    "Beslissing": [{"soort":"gevangenisstraf","aantal":6,"eenheid":"jaar"}]
}
\end{verbatim}

\begin{table}
\caption{Issues found in the court judgments}
\label{tab:issues}
	\begin{tabular}{l l @{} c @{~} p{85mm} }
\hline\noalign{\smallskip}
ECLI:NL:\dots		&date 	&cat.	&issue \\
\noalign{\smallskip}\hline\noalign{\smallskip}
RBUTR:2012:BX2092&19/07/2020	&A	&IMEI number in court judgment \\
RBNHO:2019:5662	&30/03/2020	&A	&address of offender in court judgment \\
RBNHO:2019:5823	&03/30/2020	&A	&address of offender in court judgment \\
RBNHO:2019:8774	&30/03/2020	&A	&address of offender in court judgment \\
RBDHA:2017:15274&03/03/2020	&A	&bitcoin addresses in court judgment \\
RBDHA:2020:2473	&11/08/2020	&A	&license plate in court judgment \\
RBAMS:2020:5257	&20/01/2021	&A	&name of offender in court judgment \\
RBMNE:2017:4673	&15/12/2020	&A	&name of offender in court judgment \\
RBOBR:2016:3026	&14/09/2020	&A	&name of offender in court judgment \\
RBAMS:2018:7305	&03/03/2020	&A	&passport number in court judgment \\
RBDHA:2017:15272&14/08/2020	&A	&telephone number in court judgment \\
RBOVE:2015:2282	&07/19/2020	&A	&telephone numbers in court judgment \\
RBAMS:2019:4341	&24/12/2020	&S	&year of birth of the offender \textit{1900} is wrong \\
RBLIM:2015:9808	&19/12/2020	&S	&content indication of another case included \\
RBZWB:2015:5192	&24/09/2020	&S	&unit is missing from sentence \textit{juvenile detention of 157, of which 90 conditional} \\
GHDHA:2019:2364	&01/09/2020	&S	&number and word inconsistent in penalty \textit{160 (one hundred and eighty)} \\
RBOVE:2019:1363	&02/09/2020	&S	&number and word inconsistent in penalty \textit{160 (one hundred and eighty)} \\
RBAMS:2016:9784	&28/08/2020	&S	&number and word inconsistent in sentence \\
RBAMS:2019:9707	&31/08/2020	&S	&number and word inconsistent in sentence \\
RBROT:2018:1383	&21/08/2020	&S	&number and word inconsistent in sentence \textit{35 (thirty) months} \\
RBNHO:2015:5447	&20/11/2020	&V	&paragraph of legal article is missing \textit{10a} in legal basis \\
RBROT:2018:3004	&21/11/2020	&V	&non-existent legal article \textit{10310} in legal basis\\
RBMNE:2015:3302	&14/07/2020	&V	&wrong article \textit{240bis} in legal basis \\
RBMNE:2017:1391	&19/11/2020	&V	&wrong article \textit{273 instead of 273f} in legal basis \\
RBAMS:2017:9089	&14/07/2020	&V	&wrong article \textit{240bis instead of 420bis} in legal basis \\
RBAMS:2015:7499	&21/11/2020	&V	&law is lacking in legal basis \\
RBAMS:2020:410	&28/11/2020	&V	&law is lacking in legal basis \\
RBDHA:2020:10159&28/11/2020	&V	&law is lacking in legal basis \\
RBLIM:2015:3769	&29/10/2020	&V	&law is lacking in legal basis \\
RBLIM:2015:6347	&29/10/2020	&V	&law is lacking in legal basis \\
RBLIM:2018:2963	&21/11/2020	&V	&law is lacking in legal basis \\
RBNNE:2020:1618	&28/11/2020	&V	&law is lacking in legal basis \\
RBROT:2017:6177	&21/11/2020	&V	&law is lacking in legal basis \\
RBROT:2019:7637	&10/30/2020	&V	&law is lacking in legal basis \\
RBAMS:2018:8767	&29/10/2020	&V	&article of law is missing in legal basis \\
RBNHO:2020:1960	&28/11/2020	&V	&article of law is missing in legal basis \\
RBROT:2020:6265	&28/11/2020	&V	&article of law is missing in legal basis \\
RBMNE:2019:5886	&20/11/2020	&V	&legal basis is missing \\
RBNNE:2017:2881	&10/12/2020	&V	&legal basis is missing \\
RBLIM:2015:6347	&21/11/2020	&V	&legal basis is missing \\
\noalign{\smallskip}\hline
\end{tabular}
\end{table}

\section{Appendix: Problems in Published judgments}
\label{app:issues}
During the development of the coding program, we discovered and reported 39 non-trivial problems in the published court judgments, as shown in Table~\ref{tab:issues}.
The anonymization (category A) problems have now been solved.
Category S concerns non-trivial problems with spelling and category V concerns problems with legal basis.

\end{document}